\documentstyle[aps,prl]{revtex}
\begin{document}
\title{How to be master of your domains}
\author{J.R. Anglin}
\address{Harvard-MIT\ Center for Ultracold Atoms\\
janglin@mit.edu}
\maketitle

\begin{abstract}
{\bf Abstract:} Prepare a two-species BEC in a perfectly phase-mixed state.
\ By applying Rabi drives, one can tune the range of wavelengths of
phase-separating excitations that are dynamically unstable. \ Yada yada
yada, this determines the characteristic sizes of the eventual spin domains.
\ The trapping potential is neglected because it makes life hard, but of
course this is a terrible approximation, and the results are of only
inspirational value. \ Since this is thus a rather trivial calculation, the
very modest increase in kudos that might accrue from journal publication
does not seem to outweigh the grief of having to change the title. \ This
will therefore be an ArXiv exclusive. \ 
\end{abstract}

\bigskip 

\bigskip 

Consider a two-species condensate held in a homogeneous trap, and subjected
to a Rabi drive. \ Homogeneous stationary states (there are always at least
two solutions) satisfy the time-independent Gross-Pitaevskii equation 
\begin{eqnarray}
\mu _{1}\psi _{1} &=&Ae^{-i\Omega t}e^{i\theta _{L}}\psi _{2}+\left(
g_{11}\left| \psi _{1}\right| ^{2}+g_{12}\left| \psi _{2}\right| ^{2}\right)
\psi _{1}  \nonumber \\
\mu _{2}\psi _{2} &=&Ae^{i\Omega t}e^{-i\theta _{L}}\psi _{1}+\left(
g_{12}\left| \psi _{1}\right| ^{2}+g_{22}\left| \psi _{2}\right| ^{2}\right)
\psi _{2}\;.  \label{gpe}
\end{eqnarray}
Here $A$ is the driving amplitude, which initially we will take to be real
and positive; $\theta _{L}$ is the phase of the driving field; and\ $\Omega $
is its frequency. \ (Setting $A$ positive may obviously be done without loss
of generality, but it will be notationally convenient later to redefine it
so as to allow it to be negative.) \ 

Our goal is to determine the Bogoliubov spectrum for small perturbations
around the homogeneous stationary states, and then to consider which
Bogoliubov modes may be dynamically unstable. \ Such instabilities will
indicate the earliest, linear stages of spontaneous phase separation into
spin domains. \ By varying $A$, it will turn out to be possible to control
the ranges of wavelengths that are unstable. \ If we assume that the
nonlinear evolution of domain formation does not significantly alter the
`decisions' made in the linear epoch, about where each species predominates,
then control of the instability wavelengths means control of domain sizes 
\cite{SSTP}. \ (This is admittedly a {\it yada theorem \cite{SSTP2,yada}};
but it is probably not too far wrong.)

From the viewpoint of BEC theory, it is obviously desirable to incorporate
realistic features like an inhomogeneous trapping potential, and accurate
values for the independent components of the interaction co-efficients $%
g_{ij}$. \ To get a feel for the basic concept, however, let us consider the
tractable case of no potential, as in (\ref{gpe}), and also stipulate $%
g_{11}=g_{22}\equiv g$. \ \ It will then be a nice notation to define $%
g_{12}\equiv g\Delta $. \ We then have the two solutions 
\begin{eqnarray}
\psi _{1} &=&\;\;\sqrt{\rho }e^{i(\theta _{0}+\theta _{L}/2)}e^{-i(\mu
+\Omega /2)t}  \nonumber \\
\psi _{2} &=&\pm \sqrt{\rho }e^{i(\theta _{0}-\theta _{L}/2)}e^{-i(\mu
-\Omega /2)t}  \nonumber \\
\mu  &=&\rho (g+g_{12})\pm A\equiv \frac{\hbar }{2M\xi ^{2}}\left[ 1+\Delta
/4+\alpha /2\right] \;,  \label{branches}
\end{eqnarray}
where in the last step we introduce another notation that includes the
healing length: $g_{12}\equiv g(1+\Delta )$, $A\equiv \pm \alpha g\rho $,
and $2g\rho \equiv $ $\hbar /(2M\xi ^{2})$ for $M$\ the atomic mass.\ In
these solutions the densities $\rho $ of each of the two condensates are
equal and constant, but their phases may differ by the drive phase, or by
the drive phase plus $\pi $. \ There are thus two stationary states to
consider; and at least to a good first approximation, either of them may be
prepared at will, with an appropriate initial Rabi pulse. \ (A caveat: it
will ultimately be an important question how much dynamical instabilities
may limit the accuracy of the initial state preparation.)\ \ So we have two
Bogoliubov problems to solve, depending on the branch of the $\pm $ in (\ref
{branches}). \ Since the positive parameter $A$ already appears in our
solutions, however, we have made things more compact by absorbing the $\pm $
into $\alpha $, which can thus be positive or negative. \ So from now on we
will discuss only one set of Bogoliubov modes, whose frequencies will depend
on our arbitrary real control parameter $\alpha $. \ 

If for $j=1,2$\ we write 
\[
\psi _{j}\rightarrow \psi _{j}\left[ 1+e^{i({\bf k}\cdot {\bf x}-\omega
t)}(r_{j{\bf k}}+s_{j{\bf k}})+e^{-i({\bf k}\cdot {\bf x}-\omega _{k}t)}(r_{j%
{\bf k}}^{\ast }-s_{j{\bf k}}^{\ast })\right] 
\]
and then linearize the time-dependent Gross-Pitaevskii equation in $r_{j{\bf %
k}}$ and $s_{j{\bf k}}$, we obtain the Bogoliubov equations 
\begin{equation}
\frac{\omega _{k}}{2g\rho }\left( 
\begin{array}{c}
r_{1{\bf k}} \\ 
r_{2{\bf k}} \\ 
s_{1{\bf k}} \\ 
s_{2{\bf k}}
\end{array}
\right) =\left( 
\begin{array}{cccc}
0 & 0 & k^{2}\xi ^{2}-\alpha /2 & \alpha /2 \\ 
0 & 0 & \alpha /2 & k^{2}\xi ^{2}-\alpha /2 \\ 
k^{2}\xi ^{2}-\alpha /2+1 & \alpha /2+1+\Delta & 0 & 0 \\ 
\alpha /2+1+\Delta & k^{2}\xi ^{2}-\alpha /2+1 & 0 & 0
\end{array}
\right) \;\left( 
\begin{array}{c}
r_{1{\bf k}} \\ 
r_{2{\bf k}} \\ 
s_{1{\bf k}} \\ 
s_{2{\bf k}}
\end{array}
\right) \;.
\end{equation}
This is trivially equivalent to 
\begin{equation}
\frac{\omega _{k}}{2g\rho }\left( 
\begin{array}{c}
r_{1{\bf k}}+r_{2{\bf k}} \\ 
s_{1{\bf k}}+s_{2{\bf k}} \\ 
r_{1{\bf k}}-r_{2{\bf k}} \\ 
s_{1{\bf k}}-s_{2{\bf k}}
\end{array}
\right) =\left( 
\begin{array}{cccc}
0 & k^{2}\xi ^{2} & 0 & 0 \\ 
k^{2}\xi ^{2}+2+\Delta & 0 & 0 & 0 \\ 
0 & 0 & 0 & k^{2}\xi ^{2}-\alpha \\ 
0 & 0 & k^{2}\xi ^{2}-\alpha -\Delta & 0
\end{array}
\right) \;\left( 
\begin{array}{c}
r_{1{\bf k}}+r_{2{\bf k}} \\ 
s_{1{\bf k}}+s_{2{\bf k}} \\ 
r_{1{\bf k}}-r_{2{\bf k}} \\ 
s_{1{\bf k}}-s_{2{\bf k}}
\end{array}
\right) \;.
\end{equation}
Hence we have a `sonic branch' of excitations, in which the two species
oscillate in phase, and in which the frequencies are always real (and in
fact it can be shown, positive) as long as $\Delta $ is not too negative. \
For all condensates prepared so far, $\left| \Delta \right| \ll 1$, so these
modes are not very interesting.

On the other hand, we have an `optical branch' in which the two species
oscillate out of phase, and here is the physics we want to see. \ The
frequency of an optical mode of wave number ${\bf k}$ is, for $k\equiv
\left| {\bf k}\right| $, 
\begin{equation}
\omega _{k}=2g\rho \sqrt{\left( k^{2}\xi ^{2}-\alpha \right) \left( k^{2}\xi
^{2}-\alpha -\Delta \right) }\;,
\end{equation}
where the sign of any real $\omega _{k}$\ does not concern us here. \ (A
negative real frequency for a positive norm mode means only energetic
instability, and we are interested in dynamical instabilities, for which the
Bogoliubov norm is zero.) \ What matters to us is that $\omega _{k}$\ will
be imaginary, indicating a dynamical instability, if 
\begin{eqnarray}
\alpha +\Delta <k^{2}\xi ^{2}<\alpha \; &,&\;\;\Delta <0  \nonumber \\
\alpha <k^{2}\xi ^{2}<\alpha +\Delta \; &,&\;\;\Delta >0\;.
\end{eqnarray}
\ So with negative $\Delta $, the two species intrinsically prefer to stay
mixed; but by applying a positive $\alpha $ we can destabilize them toward
phase separation. \ Positive $\Delta $,\ on the other hand, means that the
condensates would intrinsically prefer to separate, with domain sizes no
smaller than $\xi \Delta ^{-1/2}$ (assuming that the minimum domain size is
given by the maximum dynamically unstable $k$). \ \ By applying negative $%
\alpha $ we can enlarge this minimum domain size; if we make it larger than
the sample size, then we effectively frustrate the phase separation. \ 

For either sign of $\Delta $, we can turn on phase separation by applying
positive $\alpha $. \ We can make the minimum domain size smaller; and for
large enough $\alpha $, we can even introduce a maximum domain size. \ Hence
we can truly be masters of our domains.


\begin{references}
\bibitem{yada}  The assumption that the initial linear epoch controls the
final domains is not obviously unreasonable; but it is not obviously valid,
either. \ Even within Gross-Pitaevskii, the assumption needs to be checked
by following the full, nonlinear evolution, with dissipation. \ Moreover, in
the linear epoch there will typically be several competing unstable modes,
and excitations of all of them will grow in quantum superposition, until
mean field and Bogoliubov theory break down (after a time which grows only
logarithmically in the particle number $N$). \ This then raises a deep and
open problem, hitherto raised mainly in cosmology: in a nonlinear quantum
field theory, how do quantum fluctuations at early times `mature' into
classical inhomogeneities at late times? \ So the experiment would be very
interesting from a fundamental point of view.
\end{references}
\end{document}